\documentclass[aps,twocolumn,showpacs,preprintnumbers,amsmath,amssymb,superscriptaddress,prb]{revtex4-1}

\usepackage{graphicx}
\usepackage{epstopdf}
\usepackage{dcolumn}
\usepackage{bm}

\begin{document}

\preprint{}

\title{The role of magnetic anisotropy in spin filter junctions}

\author{R. V. Chopdekar}
\email[Electronic address: ]{rvc2@cornell.edu}
\altaffiliation[Current affiliation: ]{Laboratory for Micro- and Nanotechnology, Paul Scherrer Institut,
CH-5232 Villigen PSI, Switzerland}
\affiliation{Department of Materials Science and Engineering, University of California, Berkeley, Berkeley, CA 94720}
\affiliation{School of Applied and Engineering Physics, Cornell University, Ithaca, NY 14853}
\author{B. B. Nelson-Cheeseman}
\affiliation{Department of Materials Science and Engineering, University of California, Berkeley, Berkeley, CA 94720}
\author{M. Liberati}
\affiliation{Advanced Light Source, Lawrence Berkeley National Laboratory, Berkeley, CA 94720}
\author{E. Arenholz}
\affiliation{Advanced Light Source, Lawrence Berkeley National Laboratory, Berkeley, CA 94720}
\author{Y. Suzuki}
\affiliation{Department of Materials Science and Engineering, University of California, Berkeley, Berkeley, CA 94720}
\affiliation{Materials Science Division, Lawrence Berkeley National Laboratory, Berkeley, CA 94720}

\date{\today}

\begin{abstract}
We have fabricated oxide based spin filter junctions in which we demonstrate that magnetic anisotropy can be used to tune the transport behavior of spin filter junctions. Until recently, spin filters have been largely comprised of polycrystalline materials where the spin filter barrier layer and one of the electrodes are ferromagnetic. These spin filter junctions have relied on the weak magnetic coupling between one ferromagnetic electrode and a barrier layer or the insertion of a nonmagnetic insulating layer in between the spin filter barrier and electrode. We have demonstrated spin filtering behavior in La$_{0.7}$Sr$_{0.3}$MnO$_{3}$/chromite/Fe$_{3}$O$_{4}$ junctions without nonmagnetic spacer layers where the interface anisotropy plays a significant role in determining transport behavior. Detailed studies of chemical and magnetic structure at the interfaces indicate that abrupt changes in magnetic anisotropy across the non-isostructural interface is the cause of the significant suppression of junction magnetoresistance in junctions with MnCr$_{2}$O$_{4}$ barrier layers.
\end{abstract}


\maketitle

\section{\label{sec:level1}Introduction}
Spin polarized devices such as magnetic tunnel junctions have been recognized as potential building blocks for a new type of spin based electronics in recent years. While magnetic tunnel junctions, which are composed of two ferromagnetic electrodes sandwiching an insulating barrier, were first conceived in 1975 by Julliere,\cite{Julliere}  it was not until the 1990s that significant junction magnetoresistance (JMR) was demonstrated in magnetic tunnel junctions at room temperature\cite{PhysRevLett.74.3273} and it was realized that transport through these structures is extremely sensitive to the interface scattering and spin polarized interface density of states of the electrode.\cite{PhysRevB.70.054416} Briefly, in magnetic tunnel junctions it is the relative orientation of the electrode magnetization that determines whether the junction exhibits a high or low resistance state with the JMR being defined as the fractional change of resistance between these two states. It was not until recently, however, that the importance of understanding the role of the barrier layer in the tunneling process was recognized in experimental and theoretical studies of magnetic tunnel junctions with MgO barriers.\cite{ PhysRevB.63.054416} In these junctions, the symmetries of the propagating states in the electrodes and the evanescent states in the barrier, interface resonance states as well as the details of the chemical bonding between the atoms in the electrodes and barrier were recognized to be important factors in describing the spin transport.

Another important class of spin polarized devices is a spin filter device in which one electrode and the barrier layer are ferromagnetic; the relative orientation of the magnetization in the two layers again determines whether the device is in a high or low resistance state. In such devices, the ferromagnetic barrier layer has spin filtering functionality and has often been simply described as a finite potential barrier whose height depends on the spin polarization of the carrier. However it is clear that interaction between the carriers and the barrier make spin transport more complicated. In any case, effective spin filtering can occur when the two ferromagnetic layers are magnetically decoupled so that one can obtain a significant difference in resistance between the parallel low resistance state and the anti-parallel high resistance state. This magnetic decoupling had, up until recently, only been realized in polycrystalline spin filter junctions with and without a nonmagnetic layer separating the two ferromagnetic layers.\cite{leclairAPL625, PhysRevLett.99.016602} 

Recently, however, spin-filtering behavior has been observed in epitaxial oxide junctions.\cite{PhysRevB.76.220410, RamosAPL, LudersPRB} Although some of these studies are based on junctions with a nonmagnetic spacer layer between the ferromagnetic spin filter barrier layer and the ferromagnetic electrode, others demonstrated that spin-filtering behavior can be obtained without this nonmagnetic spacer. For example, we have studied junctions composed of one cubic perovskite structure  La$_{0.7}$Sr$_{0.3}$MnO$_{3}$ (LSMO) electrode, a spinel structure barrier layer and spinel structure electrode. \cite{ PhysRevB.76.220410} The weak magnetic decoupling occurs at the interface of the ferromagnetic perovskite electrode and ferrimagnetic spinel barrier layer due to magnetic frustration. In these junctions, a ferrimagnetic Fe$_3$O$_4$ electrode was used as it was strongly coupled to the barrier layer and its magnetization provided a handle with which to magnetically switch the barrier layer. To date, it is unclear how there could be little or no magnetic coupling between adjacent ferromagnetic epitaxial layers. In order to understand the weak magnetic coupling, a detailed study of the magnetism at this interface and the role of magnetic anisotropy and magnetic frustration in determining the spin filtering behavior is necessary. 

In this paper, we demonstrate that two adjacent ferromagnetic layers of a spin filter junction can be weakly magnetically coupled. The magnetic coupling at this interface, and hence the magnetotransport in the spin filter junctions, is largely determined by the magnetic anisotropy at the interface. We have fabricated LSMO/ chromite/ Fe$_3$O$_4$ junctions where the chromite barrier layer, either CoCr$_2$O$_4$ (CCO) or MnCr$_2$O$_4$ (MCO), is isostructural with Fe$_3$O$_4$. Although both chromite compounds form a normal spinel with all Cr$^{3+}$ ions in the octahedral sites, the magnetic anisotropy of the two compounds are opposite in sign and thus give rise to junction magnetoresistance values over an order of magnitude higher in CCO junctions compared to MCO junctions. Detailed studies of chemical and magnetic structure at the interfaces in both types of junctions indicate that abrupt changes in magnetic anisotropy across the non-isostructural interface is the cause of the significant suppression of JMR in MCO junctions. The angular dependence of the junction magnetoresistance highlights the consequences of changes in interface anisotropy. Therefore magnetic anisotropy provides a means by which we can control magnetic coupling and tune junction behavior.  

\section{Experiment}

Both LSMO and Fe$_3$O$_4$ have been shown to be highly spin polarized and therefore are good candidates for magnetic tunnel junctions.\cite{nohAPL233,PhysRevB.65.064417,Huang2002261} The lattice of LSMO can be described in terms of a pseudocubic unit cell with 3.87$\AA$ on a side while Fe$_3$O$_4$ forms a cubic spinel with 8.396$\AA$ on a side. The spinel barrier layer has been chosen to be CCO or MCO which have Curie temperatures of 95K or 45K, respectively. CCO and MCO have lattice parameters of 8.333$\AA$ and 8.437$\AA$ respectively and are well matched to the Fe$_3$O$_4$. 

The trilayers of LSMO/CCO/Fe$_3$O$_4$ and LSMO/MCO/Fe$_3$O$_4$ were synthesized by pulsed laser deposition on (110) oriented SrTiO$_3$ (STO) substrates supplied by Crystec GmbH. Commercial sintered powder targets of stoichiometric single-phase oxides were used for ablation at an energy density of 1-1.5 J/cm$^2$.   Deposition parameters for the layers are as follows:  LSMO in 320 mTorr of O$_2$ at 700 $^{\circ}$C; Fe$_3$O$_4$ in a vacuum of better than 4x10$^{-6}$ Torr at 450 $^{\circ}$C; MnCr$_2$O$_4$ and CoCr$_2$O$_4$ in 25 mTorr of O$_2$ at 600 $^{\circ}$C. Thicknesses of the LSMO and Fe$_3$O$_4$ electrodes were approximately 30-50 nm while the chromite barriers were 2-4 nm thick. Following thin film growth, one half of twin samples were characterized for coercive fields and morphology while the other half were fabricated into junctions between 4x4 $\mu$m$^2$ and 40x40 $\mu$m$^2$ in area.  The junctions were fabricated by conventional photolithography and Ar ion milling. In addition, bilayer samples of (110)STO/LSMO/chromite and (110)STO/chromite/Fe$_3$O$_4$ were synthesized in order to probe the non-isostructural and isostructural interfaces respectively using element specific X-ray absorption spectroscopy (XAS) and X-ray magnetic circular dichroism (XMCD) spectroscopy. The surface sensitive nature of these probes required us to make these bilayers with a top layer thickness of less than 5 nm to ensure that we were able to probe the two types of interfaces. 

The structure of our films was characterized by X-ray diffraction on a Philips Analytical X'pert MRD diffractometer and by cross sectional transmission electron microscopy (TEM) using the Philips CM300 microscope at the National Center for Electron Microscopy in Lawrence Berkeley National Laboratory. Bulk magnetization measurements were performed in a Quantum Design MPMS 5XL magnetometer and resistivity measurements were performed in a modified Quantum Design Physical Property Measurement System. XAS and XMCD experiments in total electron yield (TEY) mode were performed at beamlines 4.0.2 and 6.3.1 of the Advanced Light Source (ALS) at Lawrence Berkeley National Laboratory. Spectroscopy experiments were performed with the sample surface normal 60$^{\circ}$ inclined from the x-ray beam from 15 K - 300 K in fields of up to 0.8 T.

\section{Structure}

Structural characterization in the form of four circle X-ray diffraction and transmission electron microscopy were performed. Phase-contrast TEM imaging shows that chromite-ferrite interfaces show excellent registry with minimal defects (not shown). Good registry between perovskite and chromite film layers can be obtained with little disorder at the non-isostructural interface.  A combination of high temperature and highly energetic species during growth make it difficult to avoid interdiffusion of chemical species. We have demonstrated in a previous study that nanoscale cation migration does occur at isostructural interfaces and that it induces room temperature ferromagnetism in the chromite.\cite{ChopdekarIETM,ChopdekarCCOMCO}  In order to correlate the structure with magnetism, we have used XAS and XMCD to probe the chemical and magnetic structure in an element specific manner at both interfaces. 

\section{Magnetism}

An understanding of the magnetism in the LSMO, Fe$_3$O$_4$ and chromite layers as well as at their interfaces is crucial in determining the dominant mechanism in the transport of junctions composed of these materials. Through a combination of bulk film magnetometry and surface-sensitive XMCD, we have developed a complete picture of the magnetism in these junction trilayers. 

\begin{figure}
\includegraphics[width=8 cm]{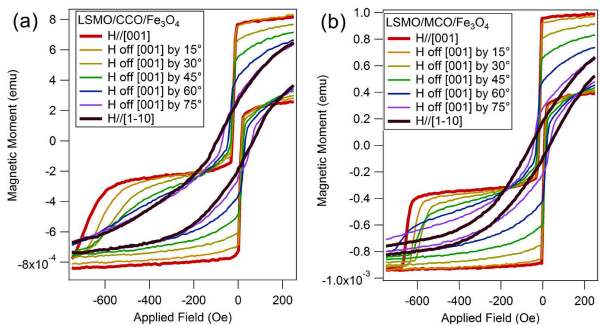}
\caption{\label{fig1}(Color online) Major magnetic hysteresis loops for unpatterned trilayers with CCO barrier (left) or MCO barrier (right). }
\end{figure}

Bulk magnetization measurements of the trilayers reveal a magnetically easy axis along the in-plane $[001]$ direction and a hard axis perpendicular in the  $[1\bar{1}0]$ direction as shown in Figure \ref{fig1} for both types of junctions. Despite small differences in the plots, we observe distinct parallel and anti-parallel electrode magnetization states along the $[001]$ direction. We note that for CCO films on (110) STO substrates, we observe uniaxial magnetic anisotropy with a [001] easy axis and a hard $[1\bar{1}0]$ axis  \cite{ChopdekarCCOMCO}. However for MCO films on [110] STO substrates, the sign of the magnetic anisotropy is reversed with the easy axis now being along the  $[1\bar{1}0]$ axis and the hard axis along the [001] axis. Both LSMO and Fe$_3$O$_4$ films on (110)STO show uniaxial anisotropy with an easy [001] axis. When the LSMO, Fe$_3$O$_4$ and chromite layers are incorporated into a trilayer, the [001] direction remains the easy direction.  Because the magnetic signal from the chromite barrier layer is so small, we cannot clearly probe the chromite magnetism in the heterostructures via SQUID magnetometry. In order to study the magnetism of the chromite layer and at its two interfaces, we used XMCD to probe the two bilayer samples described above. 

\begin{figure}
\includegraphics[width=8 cm]{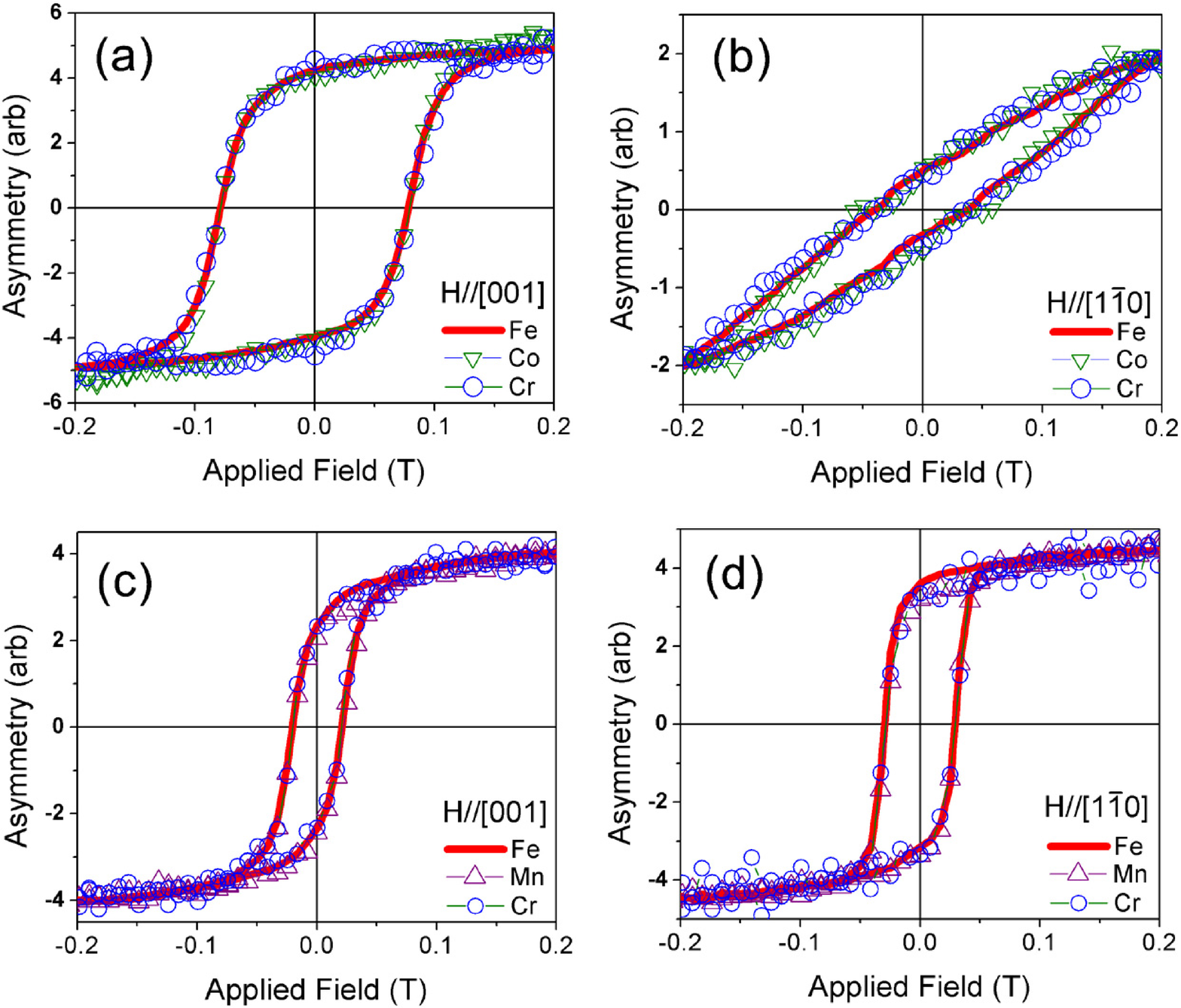}
\caption{\label{fig2}(Color online)  Room temperature element-specific hysteresis loops for an Fe$_3$O$_4$/CCO/STO sample measured with magnetic field along the (a) $[001]$ or (b)  $[1\bar{1}0]$ in-plane direction, and an Fe$_3$O$_4$/MCO/STO sample along the (c) $[001]$ or (d)  $[1\bar{1}0]$ in-plane direction.
}
\end{figure}

Let us first consider the isostructural chromite/Fe$_3$O$_4$ interfaces. Magnetic characterization using XMCD provides us with magnetic moment as a function of magnetic field and temperature in an element specific manner, thus enabling us to probe the coupling among magnetic species across the interfaces. At low temperatures (below the chromite T$_{c}$) one may expect that the ferrimagnetic chromite layers strongly exchange couple to the Fe$_3$O$_4$, but it is less clear as to the nature of the coupling above the chromite T$_{c}$.  Room temperature element-specific hysteresis loops at the Fe L$_3$ edge along the $[001]$ and $[1\bar{1}0]$ directions are shown in the solid lines of Figure \ref{fig2}. Coincident loops of Cr, Co, and Fe and Cr, Mn and Fe (Figure \ref{fig2} (a)-(d)) confirm that the interface chromite layer is coupled strongly to the Fe$_3$O$_4$ layer even at room temperature. The hysteresis loops indicate that the presence of Co and Mn have marked effect on the anisotropy and coercivity of the adjacent Fe$_3$O$_4$ cap layer, even though it is the Cr that interdiffuses more strongly into the Fe$_3$O$_4$ according to EELS data from our previous work.\cite{ChopdekarCCOMCO}. The Fe$_3$O$_4$ in our CCO/Fe$_3$O$_4$ bilayers show an increase in coercive field to approximately 1000 Oe along the $[001]$ direction, and the sample could not be saturated even in 2000 Oe along the $[1\bar{1}0]$ direction.   The Fe$_3$O$_4$  in MCO/Fe$_3$O$_4$ bilayers show coercive fields of  approximately 500 Oe but with an easy axis along the $[1\bar{1}0]$ in-plane direction.  The coercivity and anisotropy behavior in our samples matches the behavior in cobalt and manganese ferrite, and thus the Cr at the interface does not have a large influence on determining the anisotropy of the room temperature interface magnetism.  

At the nonisostructural interface, we find significantly less magnetic coupling between the LSMO and chromite layers. Previously, it was found that the growth of a spinel structure material on top of a cubic rocksalt or perovskite with half the unit cell can give rise to anti-phase boundaries and misfit dislocations.\cite{PhysRevLett.79.5162,HuIETM}  These defects, along with the ferromagnetic LSMO and ferrimagnetic chromite lattices, give rise to magnetic frustration. In order to probe the magnetism of such an interface in more detail, the (110) LSMO/chromite bilayers were explored in an analogous manner to the Fe$_3$O$_4$/chromite bilayers.  

Figure 3 shows XAS and XMCD lineshapes for a SrTiO$_3$/LSMO/MCO sample. Above the MCO T$_{c}$, the XMCD lineshapes at the Mn L$_{2,3}$ edge are characteristic of octahedral Mn$^{3+}$ and Mn$^{4+}$, similar to those in a LSMO/STO sample. The corresponding XAS lineshapes show features characteristic of tetrahedral Mn$^{2+}$ in the MCO top layer, thus indicating that Mn$^{2+}$ does not contribute magnetic signal above the MCO T$_{c}$. Below the MCO T$_{c}$, the Mn XMCD lineshape becomes dominated by the magnetism in the MCO layer. 

\begin{figure}
\includegraphics[width=8 cm]{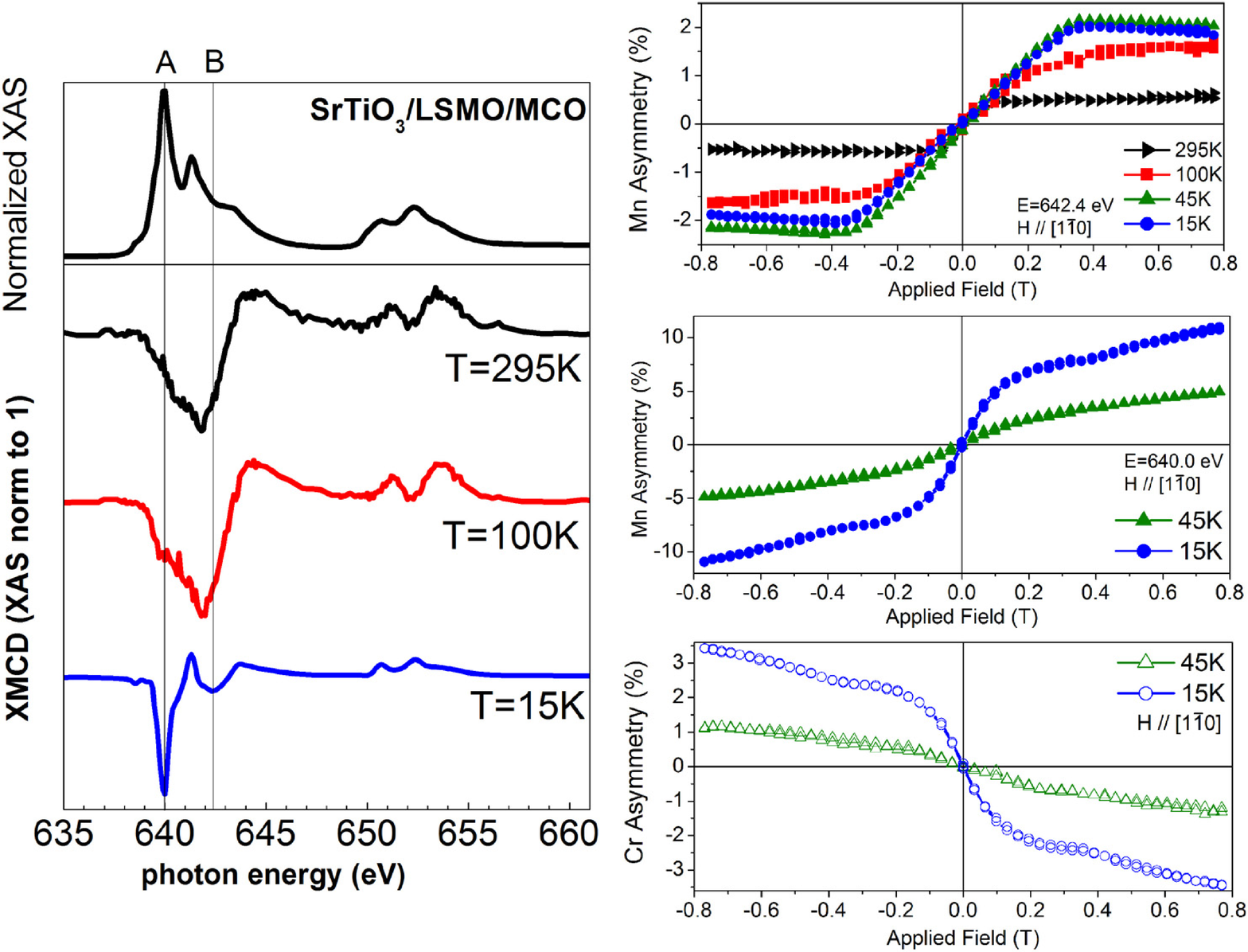}
\caption{\label{fig3}(Color online) (a) Mn L$_{2,3}$ XAS and XMCD lineshapes of an LSMO/MCO capped sample as a function of temperature, with (b)-(d) as element-specific hysteresis loops of Mn or Cr taken along the $[1\bar{1}0]$ in-plane direction.  Line A denotes E=640.0 eV, and line B denotes E=642.4 eV.}
\end{figure}

From the Mn L$_{2,3}$ lineshapes, it is clear that XMCD probes Mn in both the LSMO and MCO layers. However if we tune the photon energy to 640.0eV (line A) where we observe the maximum dichroism signal for the MCO layer or to 642.2eV (line B) where we observe the maximum dichroism signal for LSMO but close to zero dichroism for the MCO layer, we can probe the field dependence of Mn in either the MCO or LSMO layer. Figure \ref{fig3} (b) indicates that the (110) LSMO retains its uniaxial anisotropy with the magnetically hard direction along the in-plane $[1\bar{1}0]$ direction.  The small reduction in magnitude between 15 K and 45 K is an artifact due to a small positive contribution of the MCO dichroism lineshape reducing the LSMO dichroism at 642.4 eV.  Mn hysteresis loops taken at 640.0 eV and Cr hysteresis loops  along the $[1\bar{1}0]$ direction (Figure \ref{fig3} (c) and (d)) show that the MCO layer is frustrated by the LSMO underlayer and does not saturate even out to 8000 Oe, despite the $[1\bar{1}0]$ direction being the easy axis for (110) MCO single layers.  Thus the orthogonal easy axes for (110) LSMO and (110) MCO frustrate each other as is evident in the lack of saturation in the hysteresis loops.  

\begin{figure}
\includegraphics[width=8 cm]{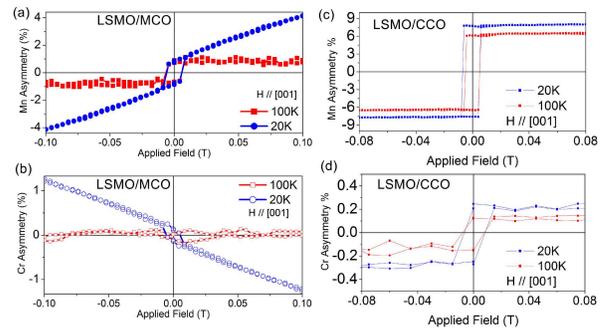}
\caption{\label{fig4}(Color online) Mn L$_{2,3}$ and Cr L$_{2,3}$ element-specific hysteresis loops of an LSMO/MCO bilayer sample ((a) and (b)) and Co L$_{2,3}$ and Cr L$_{2,3}$ loops of an LSMO/CCO bilayer sample ((c) and (d)) along the $[001]$ in-plane direction.}
\end{figure}

Similar results may be obtained from the equivalent Mn and Cr loops measured along the [001] direction (Figure \ref{fig4} (a) and (b)).  Above the MCO T$_{c}$ the Mn in the LSMO layer saturates in a field of less than 0.02 T and there is no magnetic signal from the Cr in the MCO.  When the temperature is reduced to below the T$_{c}$ of the MCO in bulk, the MCO magnetization prevents saturation of the LSMO up to fields of 0.1 T, with similar behavior seen in the Cr and Mn edge hysteresis loops (Figure 4(a) and (b)).  

For an LSMO/CCO sample (Figure \ref{fig4} (c) and (d)), the Mn in the LSMO layer switches sharply at temperatures both near and well below bulk CCO T$_{c}$.  The low Cr saturation asymmetry suggests that the Cr moment in the cap layer coupled to the LSMO layer is quite small.  This magnetic frustration for MCO and weak coupling for CCO cap layers has implications for magnetotransport as described below. 

\section{Junction Transport Behavior}

When these two types of interfaces are incorporated into a single magnetic junction, we observe markedly different magnetotransport behavior for the two types of junctions. High field JMR values on the order of -30 $\%$ were achieved by incorporating a CCO barrier layer with LSMO and Fe$_3$O$_4$ electrodes and further studies have confirmed that similar barrier layers such as FeGa$_2$O$_4$, Mg$_2$TiO$_4$, and NiMn$_2$O$_4$ can produce similarly large JMR values.\cite{alldredgeAPL} The relatively high JMR values compared to other epitaxial oxide based junctions is due in part to the use of (110) oriented LSMO in which the surface magnetization is more bulk-like than the (001) orientation.\cite{ChopdekarPRB} Despite the substantial JMR that we observe in CCO junctions, we found almost an order of magnitude smaller JMR in corresponding junctions with MCO barrier layers. A detailed investigation of the temperature and bias dependence of the JMR provides us insight into the transport mechanism and the source of this contrasting behavior.

\begin{figure}
\includegraphics[width=8 cm]{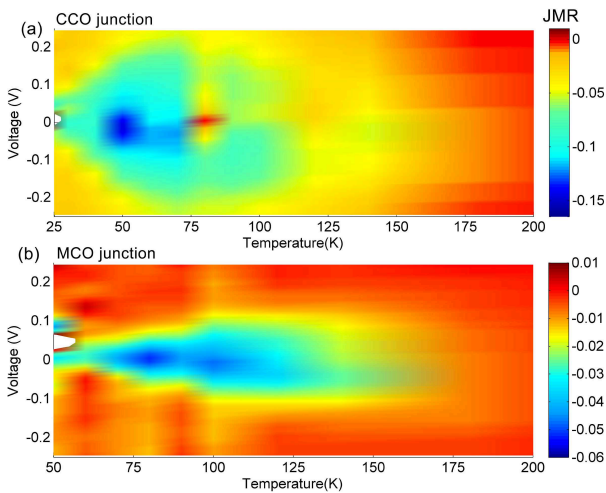}
\caption{\label{fig5}(Color online) (a) Junction magnetoresistance map as a function of bias and temperature for a device with a 4 nm CCO barrier layer (top) and a 4nm  MCO barrier layer (bottom). }
\end{figure}

The voltage and temperature dependence of the JMR can be summarized in a two-dimensional plot as shown in Figure \ref{fig5} (a) for a Fe$_3$O$_4$/4nmCCO/LSMO junction. A quick look at the plot indicates that there are three temperature regimes: T=0-70 K, T=70-175 K, and T=175-300 K.  In the lowest temperature region, the JMR decreases with decreasing temperature in contrast to the expected increase of LSMO spin polarization with decreasing temperature.  In this temperature regime, the Fe$_3$O$_4$ electrode resistance is large and increases with decreasing temperature due to the Verwey metal-insulator transition; thus the JMR is dominated by the Fe$_3$O$_4$ resistance.  

In the intermediate temperature region, the bias dependence of the JMR is asymmetric and the JMR increases with decreasing temperature.  Figure \ref{fig5} illustrates this asymmetry quite clearly for junctions with 4 nm CCO and MCO barriers.  In this temperature region, the spin polarization of the electrodes is large at low temperatures, but the asymmetric structure of the barrier/electrode interfaces produces an asymmetric conduction barrier. There have been numerous studies of magnetic tunnel junctions where asymmetries in the JMR bias dependence have been attributed to the two different interface density of states at the two electrode-barrier interfaces.\cite{Moodera1999248} In our case, it is not surprising that the isostructural and non-isostructural interfaces give rise to distinctly different density of states. We also observe a zero bias anomaly whose origin we attribute to the opening up of a charge gap in the Fe$_3$O$_4$ below the Verwey transition. The JMR minimum at 50-100 mV is consistent with observed charge gaps in Fe$_3$O$_4$.\cite{PhysRevB.58.3717}

In the highest temperature region, the magnitude of JMR is negligible and has little bias dependence. One might wonder why the spin polarization seems to decrease so much above 175 K if the Fe$_3$O$_4$ T$_{c}$ is 858K and the LSMO T$_{c}$ is 360 K . Our previous study on the temperature dependence of the magnetic coupling at the Fe$_3$O$_4$/CCO interface indicates that the magnetization of the Fe, Cr and Co sublattices decrease substantially between T=150-200K.\cite{ChopdekarCCOMCO} Thus it is expected that the spin-filtering efficiency for the exchange-coupled chromite-Fe$_3$O$_4$ bilayer also decreases substantially in this temperature region.  Additionally, temperature-dependent measurements of magnetic junctions with LSMO electrodes and nonmagnetic barrier layers have shown that the interface spin polarization is suppressed almost as much as the suppression of LSMO surface spin polarization.\cite{PhysRevB.69.052403,PhysRevLett.81.1953}  Suppression of spin polarization at both interfaces leads to a vanishingly small JMR at room temperature.

\begin{figure}
\includegraphics[width=8 cm]{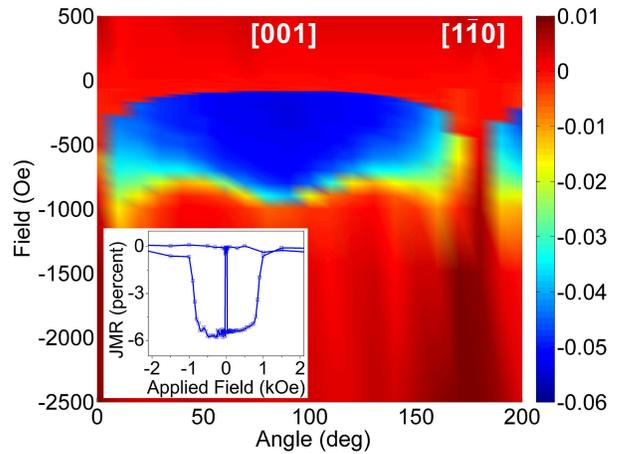}
\caption{\label{fig6}(Color online) Junction magnetoresistance map as a function of magnetic field and azimuthal angle for a 2 nm CCO based junction.  Inset: JMR hysteresis loop as a function of field along the $[001]$ in-plane direction.}
\end{figure}

\begin{figure}
\includegraphics[width=8 cm]{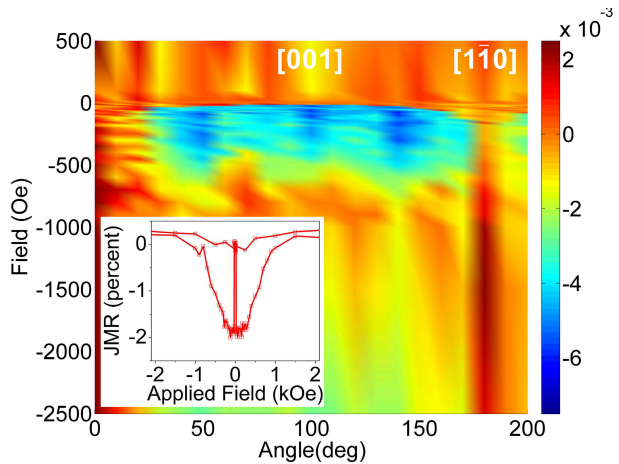}
\caption{\label{fig7}(Color online) Junction magnetoresistance map as a function of magnetic field and azimuthal angle for a 2 nm MCO based junction.  Inset: JMR hysteresis loop as a function of field along the $[001]$ in-plane direction. }
\end{figure}

JMR measurements on MCO junctions showed significantly suppressed maximum JMR values, on the order of -1 $\%$, compared with corresponding CCO junctions. In order to explain this suppression of JMR, we probed the bulk magnetic response of the trilayer as a function of magnetic field direction. For fields applied along the $[001]$ direction, both junctions exhibit well defined parallel and anti-parallel magnetic states at all temperatures. Typical JMR versus applied field curves are shown as insets to figures \ref{fig6} and \ref{fig7}. In fact, it is the MCO junction that has sharper magnetic transitions for both the LSMO and Fe$_3$O$_4$ electrodes(Figure \ref{fig1}). Therefore the suppressed JMR must be due to the differences in interface magnetic anisotropy at the LSMO/chromite interfaces. 

If indeed the interface magnetic anisotropy is the cause of the JMR suppression, the JMR in MCO and CCO junctions as a function of the applied magnetic field direction should be distinctly different. The angular dependence of the JMR is shown in Figures \ref{fig6} and \ref{fig7} for CCO and MCO junctions respectively. For each plot, the temperature is fixed at 130K and the sample is saturated at 30kOe for each in-plane angle measured. The JMR values are normalized to the zero field resistance values. The maximum JMR values for both junctions are found to be along the $[001]$ direction while the minimum values are along the $[1\bar{1}0]$  directions. However at this temperature, the maximum JMR value for the CCO junction in Figure \ref{fig6} is -6$\%$ which is an order of magnitude higher than that for the MCO junction of -0.7 $\%$. In addition, there appear additional JMR extrema along the $[1\bar{1}1]$  directions in the MCO junctions. In order to explain the angular dependence of the JMR of the MCO junctions, we need to consider possible structural modification at both interfaces. In our previous studies of the chemical and magnetic structure of chromite-Fe$_3$O$_4$ interfaces, we have found that the interfaces show long range magnetic order  of Co, Mn and Cr cations which cannot be explained in terms of the formation of interfacial MnFe$_2$O$_4$ and CoFe$_2$O$_4$ or nanoscale roughness.\cite{ChopdekarCCOMCO} If interdiffusion and disorder at the interface were the cause of the suppression of JMR in the MCO junctions, then we would expect the Mn$^{2+}$ at the interface to be not as well coupled, in comparison to Co$^{2+}$, to Fe$^{3+}$ cations near the interface. Detailed XAS and XMCD measurements reveal that Mn$^{2+}$ and Co$^{2+}$  at the interfaces are magnetized and both strongly coupled to Fe$^{3+}$. Limited interdiffusion may give rise to modification of the magnetocrystalline anisotropy constant at the interface which in turn stabilizes a local extrema in JMR along the $[1\bar{1}1]$  directions.

Therefore despite well defined parallel and anti-parallel states in the LSMO and Fe$_3$O$_4$ electrodes for both types of chromite junctions along the $[001]$ direction, it is the stabilization of CCO moments at both interfaces along the $[001]$ direction that gives rise to high JMR. The stabilization of MCO moments along the $[1\bar{1}0]$ direction gives rise to magnetic frustration and reduced JMR.

From these magnetotransport results, it is clear that the interface plays an important role in determining the spin filtering efficiency of these junctions. What is interesting to note is that strong magnetic anisotropy is induced in the chromite barrier layer even above its nominal bulk magnetic transition temperature. We had already observed proximity induced ferromagnetism in CoCr$_2$O$_4$ / Fe$_3$O$_4$ bilayers in the past.\cite{ChopdekarCCOMCO} However our present studies makes it clear that it is not the Fe$_3$O$_4$ layer that dictates the magnetic anisotropy of the chromite layer but rather the chromite/ferrite interface itself. It is this strong interface magnetic anisotropy and its coincidence (for CCO) and frustration (for MCO) with the LSMO magnetic anisotropy that dictates the transport.  

\section{Conclusion}

In summary, we have fabricated oxide-based spin filter junctions in which we have shown that the junction transport is dictated by the magnetic anisotropy at the interface between the spin filter barrier layer and each electrode. In both types of chromite junctions, the Fe$_3$O$_4$ is strongly magnetically coupled to the chromite barrier layer and is only weakly magnetically coupled to the LSMO electrode. The coincidence of the magnetically easy axes in the chromite and LSMO layers in the CCO junctions gives rise to significant junction magnetoresistance. In MCO junctions, the easy axes of the MCO and LSMO layers are perpendicular to one another, thus giving rise to magnetic frustration and suppressed junction magnetoresistance. Therefore it is clear that magnetic anisotropy at the electrode/barrier interface plays an important role in determining spin transport in this class of devices. 

\begin{acknowledgments}
The authors would like to thank Prof. Angelica Stacy for the use of her $\theta$-2$\theta$ diffractometer, Dr. Kin Man Yu from the Lawrence Berkeley National Laboratory Materials Science Division for taking RBS spectra and Franklin Wong for transmission electron microscopy on spinel heterostructures. This research is supported by the National Science Foundation (DMR 0604277).  The Advanced Light Source and the National Center for Electron Microscopy are supported by the Director, Office of Science, Office of Basic Energy Sciences, of the U.S. Department of Energy under Contract No. DE-AC02-05CH11231.
\end{acknowledgments}

\end{document}